\documentclass[prd,superscriptaddress,amsfonts,amssymb,amsmath,showpacs,onecolumn]{revtex4-2}
\usepackage{bm}
\usepackage{amsfonts}
\usepackage{latexsym}
\usepackage[latin1]{inputenc}
\usepackage{graphicx}
\usepackage{amsmath}
\usepackage{palatino}
\usepackage{ragged2e}
\usepackage{mathpazo}
\usepackage{textcomp}
\usepackage{float}
\usepackage{booktabs}
\usepackage{dcolumn}
\usepackage{multirow}
\usepackage{hyperref}
\hypersetup{colorlinks,citecolor=blue}
\usepackage{amsmath}
\usepackage{xcolor}
\usepackage{orcidlink}
\usepackage[caption=false]{subfig}
\usepackage{commath}
\def\jnl@style{\it}
\def\aaref@jnl#1{{\jnl@style#1}}

\def\aaref@jnl#1{{\jnl@style#1}}

\def\aj{\aaref@jnl{AJ}}                   
\def\apj{\aaref@jnl{ApJ}}                 
\def\apjl{\aaref@jnl{ApJ}}                
\def\apjs{\aaref@jnl{ApJS}}               
\def\apss{\aaref@jnl{Ap\&SS}}             
\def\aap{\aaref@jnl{A\&A}}                
\def\aapr{\aaref@jnl{A\&A~Rev.}}          
\def\aaps{\aaref@jnl{A\&AS}}              
\def\mnras{\aaref@jnl{Mon.~Not.~Roy.~Astron.~Soc.}}             
\def\prd{\aaref@jnl{Phys.~Rev.~D}}        
\def\prc{\aaref@jnl{Phys.~Rev.~C}}  
\def\prl{\aaref@jnl{Phys.~Rev.~Lett.}}    
\def\qjras{\aaref@jnl{QJRAS}}             
\def\skytel{\aaref@jnl{S\&T}}             
\def\ssr{\aaref@jnl{Space~Sci.~Rev.}}     
\def\zap{\aaref@jnl{ZAp}}                 
\def\nat{\aaref@jnl{Nature}}              
\def\aplett{\aaref@jnl{Astrophys.~Lett.}} 
\def\apspr{\aaref@jnl{Astrophys.~Space~Phys.~Res.}} 
\def\physrep{\aaref@jnl{Phys.~Rep.}}      
\def\physscr{\aaref@jnl{Phys.~Scr}}       
\def\commat{\aaref@jnl{Comm.~Math.~Phys.}}              
\def\science{\aaref@jnl{Science}}               
\def\cqg{\aaref@jnl{Classical Quant.~Grav.}}            
\def\jpcs{\aaref@jnl{JPCS}}                                     
\def\ijmpd{\aaref@jnl{Int.~J.~Mod.~Phys.~D}}                    
\def\grg{\aaref@jnl{Gen.~Relat.~Gravit.}}               
\def\rpp{\aaref@jnl{Rep.~Prog.~Phys.}}          
\def\npa{\aaref@jnl{Nucl.~Phys.~A}}        
\def\lrr{\aaref@jnl{Living Rev.~Rel.}}                   
\def\jcap{\aaref@jnl{J.~Cosmology Astropart.~Phys.}}    
\def\rmp{\aaref@jnl{Rev.~Mod.~Phys.}}   
\def\epjc{\aaref@jnl{Eur.~Phys.~J.~C}}

\allowdisplaybreaks[1]

\begin{document}

\color{black}       

\title{White dwarf cooling in $f(R,T)$ gravity}

\author{Snehasish Bhattacharjee\orcidlink{0000-0002-7350-7043}}
\email{snehasish@astro.ncu.edu.tw}
\affiliation{Institute of Astronomy, National Central University, 32001 Taoyuan, Taiwan}

\date{\today}

\begin{abstract}
In recent times, astounding observations of both over- and under-luminous type Ia supernovae have emerged. These peculiar observations hint not only at surpassing the Chandrasekhar limit but may also suggest potential modifications in the physical attributes of their progenitors, such as their cooling rate. This, in turn, can influence their temporal assessments and provide a compelling explanation for these intriguing observations. In this spirit, we investigate here the cooling process of white dwarfs in $f(R,T)$ gravity with the simplest model $f(R,T) = R + \lambda T$, where $\lambda$ is the model parameter.
Our modelling suggests that the cooling timescale of white dwarfs exhibits an inverse relationship with the model parameter $\lambda$, which implies that for identical initial conditions, white dwarfs in $f(R,T)$ gravity cool faster. This further unveils that in the realm of $f(R,T)$ gravity, the energy release rate for white dwarfs increases as $\lambda$ increases. Furthermore, we also report that the luminosity of the white dwarfs also depends on $\lambda$ and an upswing in $\lambda$ leads to an amplification in the luminosity, and consequently a larger white dwarf in general relativity can exhibit comparable luminosity to a smaller white dwarf in $f(R,T)$ gravity.
\end{abstract}

\maketitle

\section{ Introduction}
White dwarfs (WDs) mark the final evolutionary stage of main sequence stars with initial masses up to 10 $M_{\odot}$ \cite{shapiro} and comprise 95\%-97\% of all the observed stars in the Milky Way \cite{woosley}. These stars usually end up as WDs once all the H in the core is fused into He during the main sequence phase and later into C and O during the horizontal branch and asymptotic giant branch phases. The stars then experience the loss of their outer envelopes and what remains are dense C-O cores. These cores do not undergo any nuclear fusion and therefore cool down by radiating heat for the rest of their existence. \\
The maximum mass a WD can acquire precisely equals 1.44 $M_{\odot}$ and is known as the Chandrasekhar limit \cite{chandrasekhar}. The constraint on this mass arises from the fact that the electron degenerate pressure can only equalize the gravitational collapse up to a certain point. Any further increase in the mass of the WD either through merger or accretion results in a thermonuclear explosion due to runaway carbon ignition and can be observed as a Type Ia supernova. Type Ia supernovae serve as standard candles, exhibiting comparable absolute magnitudes when they explode. As a result, they are a valuable tool for measuring distances in the cosmos \cite{weinberg}. Remarkably, in the late 1990s, the utilization of type Ia supernovae as distance indicators led to the discovery that our Universe is experiencing accelerated expansion \cite{reiss,perlmutter}.\\
Although general relativity has been remarkably successful in explaining numerous astrophysical phenomena, it has hitherto been unable to explain, for instance, the origin of the flat galactic rotation curves and the acceleration of the Universe. In this regard, two enigmatic cosmological entities, namely dark matter and dark energy were proposed to reconsider the observations. However, the existence of these entities is yet to be confirmed observationally or experimentally \cite{fulvio}. In this spirit, geometric extensions of general relativity were introduced to provide an alternative explanation to these phenomena without requiring dark matter and dark energy \cite{boroweic,koivisto,flanagan,fay,sotiriou,bhatta1,bhatta2,bhatta3,bhatta4,dioguardi,gialamas,dimopoulos}. \\
Stellar remnants such as black holes, neutron stars, and WDs are frequently employed to set bounds on extended theories of gravity. For instance, in \cite{berti,moreas} the authors used gravitational waves from compact binaries to impose tight constraints on multiple theories of gravity, ref \cite{das1,das2,carvalho} investigated the viability of $f(R)$, and $f(R,T)$ theories of gravity in explaining the WD properties respectively (where $R$ denote the Ricci scalar and $T$ denote the trace of the stress-energy-momentum tensor), in \cite{pretel}, the authors studied the properties of neutron stars in $f(R,T)$ gravity, while in ref \cite{jain}, the authors used observational data of WDs to set bounds on Horndeski theories of gravity. \\
In this work, we shall investigate the cooling process of WDs in $f(R,T)$ theory of gravity \cite{harko}. In this modified theory of gravity, the Einstein-Hilbert action comprises a function of both the $R$ and $T$. In doing so, the modified field equations obtained after applying the least action principle provide an elegant explanation for the acceleration of the Universe \cite{harko}, and the flat galactic rotation curves \cite{zaregonbadi}. Additionally, in \cite{shabani}, the authors have shown that several $f(R,T)$ gravity models are successful in passing the solar system tests. \\
Several authors employed $f(R,T)$ theory of gravity to investigate the properties of stellar objects. For instance, in \cite{sharif}, the authors studied the stability of collapsing spherical objects comprised of isotropic matter, while in \cite{noureen,zubair}, the authors used a perturbation framework to investigate the evolution of spherical stars. \\
In the past few years, the observations of several over- and under-luminous type Ia supernovae sparked interest in the studies of WDs in several modified theories of gravity \cite{kalita2,kalita3,wojnar,kalita4,sarmah,das3,das4,garcia,althaus,corsico,ben,saltas,liu,eslam,biesiada,benvenuto}. In addition to surpassing the Chandrasekhar limit, it is also expected that WDs may undergo alterations in their physical characteristics, such as their cooling rate, consequently influencing their chronological estimations and could provide an explanation for these observations \cite{kalita}. Bearing this in mind, we shall investigate the cooling process of WDs in $f(R,T)$ gravity with the simplest model $f(R,T) = R + \lambda T$, where $\lambda$ is the model parameter. \\
The article is organized as follows: In Section \ref{sec2}, we provide an overview of $f(R,T)$ gravity and stellar structure equations. In Section \ref{sec3}, we derive a simple model for the white dwarf cooling in $f(R,T)$ gravity and in Section \ref{sec4}, we summarize and conclude the work.

\section{Overview of $f(R,T)$ gravity and stellar structure equations}\label{sec2}

The action in $f(R,T)$ can be expressed as the following \cite{harko}

\begin{equation}
    \mathcal{S} = \int \frac{\sqrt{-g}}{16 \pi G} \left[f(R,T) + \mathcal{L}_{m}\right] d^{4}x
\end{equation}
where $\mathcal{L}_{m}$ is the matter Lagrangian, $g$, the metric determinant and $G$ the Newtonian gravitational constant. In this work, we shall use natural units by setting $G=c=1$. The field equations can now be derived by varying the action with respect to the metric $g_{mn}$ as the following \cite{harko}

\begin{equation}
    f_{R}(R,T)R_{mn} + (g_{mn}\Box - \bigtriangledown_{m} \bigtriangledown_{v}) f_{R}(R,T) - \frac{1}{2}f(R,T)g_{mn} = 8 \pi T_{mn} - f_{T}(R,T)(\Theta_{mn} + T_{mn}),
\end{equation}
where $R_{mn}$ denote the Ricci tensor, $\Box = \bigtriangledown^{m}\bigtriangledown_{n}$ represent the D'Alembertian operator and $\bigtriangledown_{n}$ the covariant derivative. Additionally, 

\begin{equation}
    f_{R}(R,T) \equiv \frac{\partial f(R,T)}{\partial R}, \hspace{0.25in} f_{T}(R,T) \equiv \frac{\partial f(R,T)}{\partial T},
\end{equation}

\begin{equation}
    T_{mn} = g_{mn}\mathcal{L}_{i} - 2 \frac{\partial \mathcal{L}_{i}}{\partial g^{mn}}, \hspace{0.25in} \Theta \equiv g^{ij} \frac{\delta T_{ij}}{\delta g^{mn}}.
\end{equation}
Following \cite{harko}, we shall focus our attention on the functional form $f(R,T) = R + \lambda (T)$, where $\lambda$ represents the model parameter. Furthermore, we shall consider a spherically symmetric metric of the form
\begin{equation}
    ds^{2} = -e ^{2 \eta} dt^{2} + e ^{2 \gamma} dr^{2} +r^{2} (d\theta^{2} + \sin ^{2}\theta d \phi^{2}),
\end{equation}
where $x^{m} = (t, r , \theta, \phi)$ represent the Schwarzschild coordinates, and $\eta$, and $\gamma$ the metric potentials. \\
For delineating the WD matter, we shall be using the energy-momentum tensor for an isotropic perfect fluid with the following form
\begin{equation}
    T_{mn} = (\rho + p) u_{m} u_{n} + p g_{mn},
\end{equation}
where $\rho$ and $p$ denote the energy density and pressure of the fluid, and $u_{m}$ is the four-velocity of a comoving observer. When the WD maintain hydrostatic equilibrium, the thermodynamic quantities and metric are time-independent, and thus we arrive at the following Tolman-Oppenheimer-Volkoff (TOV) equations in $f(R,T)$ gravity \cite{tolman, volkoff,pretel}
\begin{equation}\label{dm}
    \frac{dm}{dr} = 4 \pi r^{2}\rho - \frac{r^{2}}{4} \lambda T
\end{equation}
\begin{equation}\label{dp}
    \frac{dp}{dr} = -\left(\frac{\rho + p}{1 - \frac{2m}{r}}\right) \left[4 \pi r p + \frac{m}{r^{2}} + \frac{r}{2} \left( (p + \rho) \lambda + \frac{1}{2} \lambda T\right)  \right],
\end{equation}
where $m$ is the mass of the star within radius $r$. It is clear that by setting $\lambda=0$, one readily recovers the standard TOV equations in GR \cite{tolman,volkoff}.\\

The above set of equations can be solved for an equation of state (EoS) of the matter describing the stellar content. The EoS contain information about the microstates characterizing the system and contains information about the various forces acting between the particles which are dependent on the phase transition points and temperature \cite{kalita}. For simplicity, we shall adopt a barotropic EoS of the form $P = P(\rho)$, and overlook any dependence on other thermodynamic variables such as temperature. This choice of the EoS emerges from the fact that for WDs, the Fermi temperature is much higher than its temperature. Furthermore, following \cite{kalita}, we shall restrict our analysis to non-rotating and non-magnetized WDs since they can be modelled as spherical-symmetric structures composed of degenerate electrons. Therefore, the EoS that illustrates the properties of such WDs can be expressed as the following \cite{chandrasekhar}

\begin{equation}\label{rho}
    \rho = \left(\frac{8 \pi m_{H}\mu_{e}(m_{e}c)^{3}}{3h^{3}}\right)x_{F}^{3},
\end{equation}
\begin{equation}\label{p}
    P = \frac{\pi c^{5}m_{e}^{4}}{3 h^{3}}\left[x_{F}(2x_{F}^{2}-3) \sqrt{x_{F}^{2}+1}+3 \sinh^{-1}x_{F} \right],
\end{equation}
where $m_{H}$, and $m_{e}$ represent the mass of a hydrogen atom and mass of an electron respectively, $x_{F}=p_{F}/m_{e}c$ with $p_{F}$ denote the Fermi momentum and $\mu_{e}$ the mean molecular weight per electron. Using Eqs. \ref{rho} and \ref{p} one can solve Eqs. \ref{dm} and \ref{dp} to understand the mass-radius relation of WDs in $f(R,T)$ gravity.

\section{White dwarf cooling in $f(R,T)$ gravity}\label{sec3}

We shall now investigate the cooling process of WDs in our chosen $f(R,T)$ gravity model. To make the analysis simpler, let us model the WD as a perfect blackbody that cools down by radiating away its energy without the presence of any alternate energy reservoirs. Hence, the cooling timescale depends on how the atmospheric characteristics of the WD transfer energy through itself and on the chosen theory of gravity. The cooling process of WDs has been previously investigated in GR \cite{shapiro,mestel} and in $f(R)$ gravity \cite{kalita}. \\
The equation governing the transport of radiation for the WD reads

\begin{equation}\label{dt}
    \frac{dT}{dr} = - \left[\frac{3 \kappa \rho (r)  L(r)}{16 \pi r^{2}a c T(r)^{3}} \right],
\end{equation}
where $a = 8 \pi k_{B}^{4}/ 15 c^{3}h^{3}$ represent the radiation constant with $k_{B}$ the Boltzmann constant, $T(r)$ and $L(r)$ the temperature and luminosity of the WD at radius $r$ respectively and $\kappa$ the opacity. Additionally, we can express the luminosity gradient of the WD as the following 

\begin{equation}
    \frac{dL}{dr} = 4 \pi r^{2} \rho(r) \eta(r),
\end{equation}

where $\eta(r)$ represents the power generated per unit mass of the WD. Combining Eqs. \ref{dp} and \ref{dt}, we obtain 

\begin{equation}\label{dtdp}
    \frac{\partial T}{\partial P} = \frac{3 L \left(1-\frac{2 m}{r}\right) \kappa \rho}{16 \pi  a c r^2 T^3 (\rho +p) \left(\frac{1}{2} r (0.5 \lambda(\rho - 3 p)+\lambda (\rho +p))+\frac{m}{r^2}+4 \pi  p r\right)}.
\end{equation}
To proceed further, we assume Kramer's opacity prescription with $\kappa = \alpha \rho T^{-7/2} $, where $\alpha = 10^{24}Z(1+X)cm^{2}g^{-1}$. Here, $Z$ denotes the mass-fraction of metals and $X$ is the same for hydrogen. Substituting $\kappa$ in Eq. \ref{dtdp}, we obtain

\begin{equation}\label{dpdt}
    \frac{dP}{dT} = \frac{16 \pi  a c r^2 T^{13/2} (\rho + p) \left(\frac{1}{2} r (0.5 \lambda(\rho - 3 p))+\lambda (\rho+p))+\frac{m}{r^2}+4 \pi  p r\right)}{3 \alpha  L \left(1-\frac{2 m}{r}\right) \rho ^2}.
\end{equation}
Now, following \cite{kalita}, we shall use an ideal gas EoS near the surface of the WD as follows

\begin{equation}\label{ideal}
    \rho = \frac{\mu m_{u}}{\kappa_{B} p T},
\end{equation}
where $\mu$ represents the mean molecular weight and $m_{u}$ the atomic mass unit. Furthermore, near the surface of the WD, $r \approx R$ where $R$ represents the radius of the WD and therefore $m(r \approx R) \approx M$, where $M$ represent the total mass of the WD. Thus, substituting Eq. \ref{ideal} in Eq. \ref{dpdt}, we obtain after simplification 

\begin{equation}\label{dp2}
\left( \frac{1}{P} \right)^{4} \frac{dP}{dT} \approx   \frac{4 \pi  a c \kappa_{B}^{2} (3 \lambda +16 \pi ) r^4 T^{17/2}}{3 \alpha \mu ^2 L m_{u}^{2} (2 M-R)}
\end{equation}

Integrating Eq. \ref{dp2} we obtain the following, 

\begin{equation}
    P \approx \frac{171}{2 \pi}\left[{{\frac{\alpha \mu ^2 L m_{u}^2 (R-2 M)}{a c \kappa_{B}^{2} (3 \lambda +16 \pi ) r^{4} T^{19/2}}}}\right]^{1/3}.
\end{equation}

Using Eq. \ref{ideal} the expression for $\rho$ reads,

\begin{equation}\label{rho2}
    \rho \approx \frac{171 \mu m_{u}}{2 \pi \kappa_{B}}\left[{{\frac{\alpha \mu ^2 L m_{u}^2 (R-2 M)}{a c \kappa_{B}^{2} (3 \lambda +16 \pi ) r^{4} T^{25/2}}}}\right]^{1/3}
\end{equation}

For further analysis, we shall assume $x_{F}\ll 1$ since a non-relativistic EoS is adopted for regions close to the surface. With this simplification, we use Eq. \ref{rho} to obtain $P \approx 10^{13} (\rho/ \mu_{e})^{5/3}$. Submitting this result in Eq. \ref{ideal}, we obtain the density $\rho_{*}$ near the surface $ \approx 2 \times 10^{-8} \mu_{e} T_{*}^{3/2} g cm^{-3}$. Additionally, let us presume the WD to be composed of carbon and oxygen with its surface comprising 90\% helium and 10\% metals, and therefore $X=0, Z=0.1,\mu=1.4$, and $\mu_{e} = 2$. Substituting all these results and $\rho_{*}$ into Eq. \ref{rho2} we obtain the following expression for the luminosity of the WD in $f(R,T)$ gravity 

\begin{equation}\label{luminosity}
    L [\text{erg/s}] \approx \frac{1.2\times 10^{15} (3 \lambda +16 \pi ) R^{4} T^{2}}{R - 2M}.
\end{equation}
Now, luminosity is defined as the rate of energy ($E$) emitted over a specific time interval ($t$) and can be mathematically expressed as the following 

\begin{equation}\label{energy}
    L = \frac{dE}{dt}.
\end{equation}

Substituting Eq. \ref{luminosity} into Eq. \ref{energy}, we obtain 

\begin{equation}
    \frac{1.2\times 10^{15} (3 \lambda +16 \pi ) R^{4} T^{2}}{ R - 2M} = \frac{d}{dt} \left[\frac{3 \kappa_{b} T M}{2 A m_{u}}\right],
\end{equation}
where $A$ represent the atomic mass constant. Integrating the equation from initial time $t_{ini}$ to the present epoch $t_{0}$ such that the present temperature $T_{0}$ is negligible to the initial temperature $T_{ini}$, we obtain the following expression for the cooling timescale ($\tau = t_{0} - t_{ini}$) in $f(R,T)$ gravity 

\begin{equation}\label{tau}
  \tau [\text{yrs}] \approx  \frac{2.4 \times 10^{-3} M (R-2M) \kappa_{b}}{R^{4}(3 \lambda + 16 \pi )  T_{ini}}.
\end{equation}

\begin{figure}[H]
\centering
\includegraphics[width=0.6\textwidth]{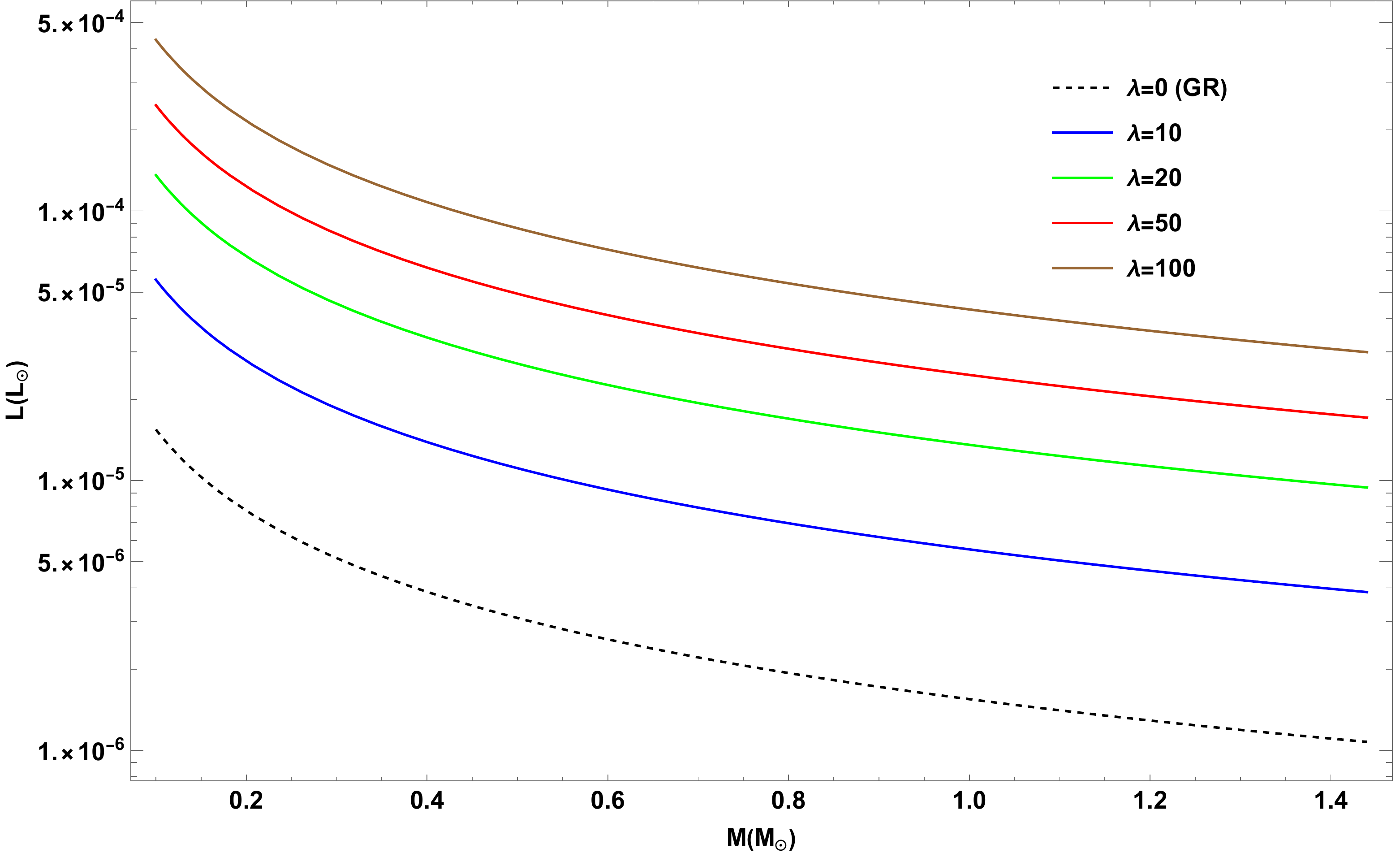}
\caption{WD luminosity ($L$) as a function of its mass ($M$) both in solar units for different $\lambda$ with radius $R=10^{8}$ cm, and with an initial temperature $T = 10^{6}$ K.  \label{lum}}
\end{figure}

We shall now investigate the behaviour of the luminosity ($L$) and the cooling timescale ($\tau$) of WDs as a function of the mass of the WD by fixing $R=10^{8}$ cm, and $T = 10^{6}$ K for different $\lambda$ which characterize deviations from GR. In Fig. \ref{lum}, we find that the luminosity decreases with an increase in the mass of the WD, and increases as the model parameter $\lambda$ increases. For the case when $\lambda = 100$, an order of magnitude enhancement in the luminosity can be seen. Furthermore, the luminosity is also an increasing function of the radius ($L \propto R^{3}$) and on the initial temperature ($L \propto T^{2}$) of the WD. Consequently, a larger WD within the realm of GR may manifest an equivalent luminosity to a relatively smaller WD in the $f(R,T)$ theory of gravity. As a result, utilizing WD luminosity could possibly define bounds on $f(R,T)$ gravity models.   \\

\begin{figure}[H]
\centering
\includegraphics[width=0.6\textwidth]{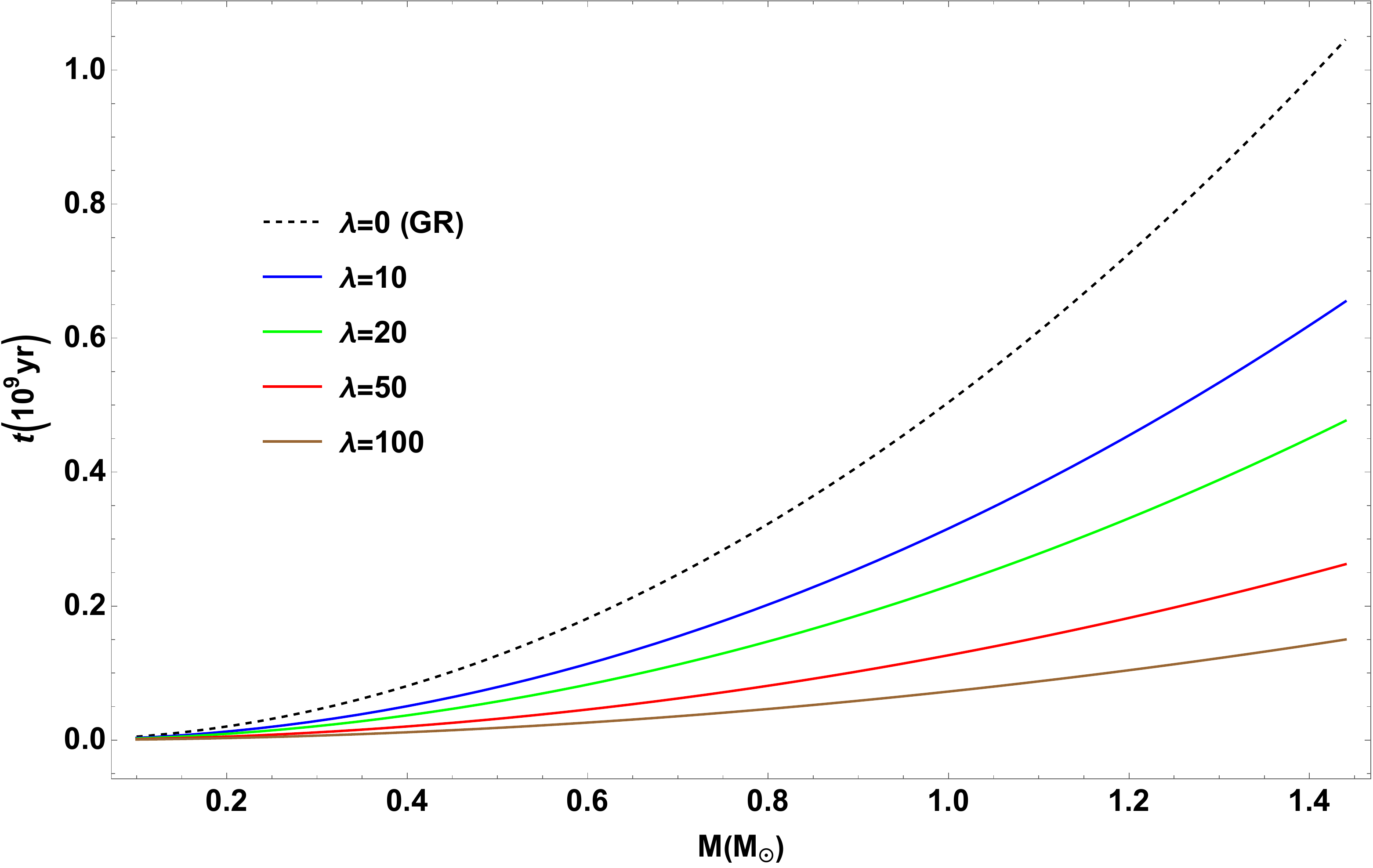}
\caption{WD cooling timescale ($\tau$) in units of $10^{9}$ yr as a function of its mass ($M$) in solar units for different $\lambda$ with radius $R=10^{8}$ cm, and with an initial temperature $T = 10^{6}$ K. \label{time}}
\end{figure}

In Fig. \ref{time}, we show the cooling timescale of WDs as a function of their mass for various $\lambda$. From Eq. \ref{tau}, we see that the cooling timescale is an increasing function of the mass ($\tau \propto M^{2}$), consequently implying massive WDs necessitate a longer duration to cool down to a lower level compared to their less massive counterparts with similar sizes and initial temperatures. Furthermore, Equation \ref{tau} elucidates that $\tau$ exhibits an inverse relationship with the model parameter $\lambda$, which implies that for identical initial conditions, WDs in $f(R,T)$ gravity cools faster. This further unveils that in the realm of $f(R,T)$ gravity, the energy release rate for WDs increases as $\lambda$ increases.

\section{Conclusions}\label{sec4}

In recent years, several over- and under-luminous type Ia supernovae have been observed which sparked an interest in the studies of WDs in several modified theories of gravity \cite{kalita2,kalita3,wojnar,kalita4,sarmah,das3,das4,garcia,althaus,corsico,ben,saltas,liu,eslam,biesiada,benvenuto}. In addition to surpassing the Chandrasekhar limit, it is believed that WDs may undergo alterations in their physical characteristics, such as their cooling rate, consequently influencing their chronological estimations and could provide an explanation for these observations \cite{kalita}. In this spirit, we investigate here the cooling process of WDs in $f(R,T)$ gravity with the simplest model $f(R,T) = R + \lambda T$, where $\lambda$ is the model parameter. \\
Our investigation makes it clear-cut that the luminosity of WDs relies not only on their masses but also on the model parameter $\lambda$. Precisely, as the mass increases, the luminosity decreases, whereas an upswing in $\lambda$ leads to an amplification in the luminosity. Furthermore, the luminosity is also an increasing function of the radius ($L \propto R^{3}$) and on the initial temperature ($L \propto T^{2}$) of the WD. Consequently, a larger WD within the realm of GR may manifest an equivalent luminosity to a relatively smaller WD in the $f(R,T)$ theory of gravity. As a result, utilizing WD luminosity could possibly define bounds on $f(R,T)$ gravity models.\\
Moreover, we also report that the cooling timescale ($\tau$) of WDs is an increasing function of the mass ($\tau \propto M^{2}$), consequently implying massive WDs necessitate a longer duration to cool down to a lower level compared to their less massive counterparts with similar sizes and initial temperatures. Furthermore, Equation \ref{tau} elucidates that $\tau$ exhibits an inverse relationship with the model parameter $\lambda$, which implies that for identical initial conditions, WDs in $f(R,T)$ gravity cools faster. This further unveils that in the realm of $f(R,T)$ gravity, the energy release rate for WDs increases as $\lambda$ increases.\\
However, we would like to emphasize that the modelling used in this work is gracefully straightforward and does not include information regarding the temporal evolution of the electron degeneracy \cite{kozak}, phase transitions \cite{auddy}, finite temperature corrections of the gas, and crystallization processes \cite{wojnar}. We aim to tackle these issues in the foreseeable future.

\section*{Acknowledgements}
SB is supported by a doctoral scholarship from the National Central University.

\end{document}